# Relation between Entropy, Diffusion and Relaxation Kinetics

**Subhajit Acharya and Biman Bagchi***

Solid State and Structural Chemistry Unit

Indian Institute of Science, Bengaluru, India

*Corresponding email: bbagchi@iisc.ac.in; profbiman@gmail.com

## Abstract

**Intermolecular correlations lower values of both diffusion and entropy. We present an analysis of the existing relations between long-time diffusion (D) and entropy. S. A recently proposed inequality, a lower bound, by Sorkin *et al.*, expresses the long-time diffusion in terms of diffusion in a reference state and the entropy difference. Such a relationship may provide a measure of intermolecular correlations. We show that for a one-dimensional rugged energy landscape, the lower bound becomes equality only if certain three-site correlations are neglected. When these correlations are included, we can derive an accurate expression that agrees with computer simulations. The strong dimensionality dependence of diffusion of a Brownian particle in a rugged energy landscape also resembles the recently proposed inequality. We show that for interacting colloids, a mode-coupling theory-type calculation of diffusion coefficient can be combined with the new inequality to estimate entropy change from long-range inter-colloid correlations. Interestingly, the rate of barrier crossing in a multidimensional potential energy surface, related to diffusion in a periodic lattice, is shown to admit a relation that depends on the entropy change in a way reminiscent of the relation between D and S.**

# I. INTRODUCTION

Relations between entropy (*S*) and diffusion (*D*) have drawn considerable attention in physics and chemistry literature ever since the seminal papers of Kauzmann and Adam, Gibbs, and DiMarzio.[1–5] It is generally accepted that diffusion tends to zero rapidly as the entropy decreases towards zero. More recently, this problem has been studied using Rosenfeld scaling, which seems to describe the relation between *S* and *D* for many liquids under ambient conditions (normal temperature and pressure). This issue also arises when analyzing random first-order (RFOT)[6–9] and mode coupling theories [10–13]. An additional aspect of this relation is the dimensionality (*d*) dependence, the origin of which is not fully clear, at least quantitatively.[14]

Entropy provides a measure of the phase space, and diffusion is a measure of the rate of exploration of the same. The fact that the two can be related is appreciated in the following fashion. Entropy can be given by the well-known Boltzmann's formula $S = k_B \ln \Omega$. On the other hand, diffusion is provided by the mean square displacement, which depends on the fluidity of the system. The fluidity depends on the accessibility of the configuration space so that the system can move from place to place in the phase space, and the tagged particle can move from position to position. It is thus understandable that diffusion and entropy could be related, although the precise form is not easily accessible. Reduction of both D and S occurs as we lower temperature or/and increase density due to the emergence of intermolecular correlations. Thus, a reliable relation can also be used to guess the extent of such correlations, as we discuss below. If a relation between diffusion and entropy can indeed be established, then it serves an important purpose because one can at least partly avoid the difficult task of calculating the friction coefficient.

Several research papers have attempted to establish a relation between the diffusion (D) of a tagged particle and the entropy of the system (S). The relation is given in the form of an inequality. In a recent paper, Sorkin *et al.*[15] derived a potentially highly useful bound on diffusion (D)-entropy (S) scaling. The relation is given in the form of an inequality [Eq.(1)]

$$D\tau \geq D_S \tau_0 \exp\left(\frac{2}{d}\Delta S\right) \quad (1)$$

Where $d$ is the dimensionality, $D_S$ is the short-time diffusion and $\Delta S$ is the entropy difference $\Delta S = S - S_0$, where $S_0$ is the per particle entropy of the reference state, and S denotes the per particle entropy of the system, as described in Ref.[15]. The relaxation times $\tau$ and $\tau_0$ characterize short-time relaxation in the real and the reference system. Eq.(1) above is Eq.5a of Sorkin et al.[15] A more restrictive expression was given in Eq.5b, where the reference state is that of the ideal gas. However, Eq.(1) is rather restrictive and less valuable because of the involvement of the relaxation times $\tau$ and $\tau_0$, which are not known precisely. Often $\Delta S$ is negative (as in Eq.5b), *then the long-time diffusion is expected to be larger than the short-time diffusion,* which, however, has not been observed in most studies, like in colloid solutions, and motion in rugged energy landscape under Langevin dynamics. The two-stage diffusion has also been observed in deeply supercooled liquids.

*Eq.(1) is an interesting result and could be valuable to a wide range of applications.* In particular, from the ratio of the two diffusion coefficients, we can get a measure of the entropy difference between the two states and also a measure of the emergence of correlations. We demonstrate this interesting possibility in the case of colloid solutions.

However, Eq.(1) is to be reconciled or verified by detailed microscopic calculations, which is partly the goal of this study. In the following, we present several pertinent calculations and theoretical analyses and make certain new advances that connect with the results reported in Ref.[15].

## II. DIFFUSION IN A RUGGED ENERGY LANDSCAPE

Several studies were earlier carried out on the diffusion of a tagged particle in a one-dimensional random, rugged energy landscape where results can be obtained rigorously both by the solution of stochastic equations and simulation methods.[16] An approximate (not-so-accurate) expression of diffusion was derived earlier by Zwanzig, which can, in turn, be used to obtain a rather neat expression reproducing diffusion-entropy scaling. Zwanzig's expression for the reduced diffusion in rugged potential can be expressed in the following form [Eq.(2)]

$$D = D_0 \exp\left[-\frac{\varepsilon^2}{(k_B T)^2}\right] \quad (2)$$

In the above expressions, $\varepsilon$ is the standard deviation (width) of the assumed Gaussian distribution in the energy distribution of the rugged landscape. It can be shown that for quenched ruggedness in a one-dimensional system, the reduced entropy (due to the ruggedness) is given exactly by $\Delta S = -\frac{\varepsilon^2}{2(k_B T)^2}$. *This leads to*

$$D = D_0 \exp[2\Delta S] \quad (3)$$

Thus, in 1d, the expression of the renormalized diffusion form *agrees* with the form obtained in Ref.[15], *with the equality sign*. As mentioned above, the entropy difference $\Delta S$ denotes the difference between the entropy of the particle in a quenched rugged energy landscape and its ideal gas *state*. In fact, this also agrees with Rosenfeld scaling.

Note that the expression for the renormalized diffusion in random rugged energy was obtained by using the expression for the mean first passage time (MFPT). It also involves the approximation *of local averaging, which removes certain pathological groupings of potential* energy surfaces. It was demonstrated in Ref.[17] that the presence of three site traps (TST) renormalizes the diffusion to the following form

$$D = D_0 \exp\left(-\frac{\varepsilon^2}{(k_B T)^2}\right) \bigg/ \left[1 + erf\left(\frac{\varepsilon}{2k_B T}\right)\right] \quad (4)$$

Incidentally, this generalized form agrees with simulation quantitatively, while the earlier form could give only semi-quantitative agreement. *It has an additional entropy dependence through the error function,* which arises from correlations among site energies in the rugged landscape. Thus, the equality is destroyed by correlations, but the inequality given by Eq.(1) seems to be violated. Note that Eq.(4) is nearly exact and non-trivial.

A precise comparison with Eq.(1) is thwarted by uncertainty posed by the two relaxation times $\tau$ and $\tau_0$.

Numerical calculations show that diffusion in a random rugged energy landscape displays a *strong dimensionality dependence*, in agreement with Eq.(1). It was demonstrated in Refs [17] that diffusion in the rugged energy landscape *increases rapidly from one dimension to two dimensions, and then the increase is gradual and finally saturates to an asymptotic result as the dimension d goes to infinity ( $d \to \infty$ ).* According to the formalism, the effective diffusion coefficient can be approximated as [Eq.(5)]

$$D = D_0 \exp\left[-\frac{\varepsilon^2}{2(k_B T)^2} - \left(\sqrt{\frac{2}{d}} - \frac{1}{\sqrt{2}}\right)\frac{\sqrt{\pi}\varepsilon}{k_B T}\right] \quad (5)$$

In 2d, Eq.(5) exactly reproduces the same derived by using the effective medium approximation. However, in other dimensions, it slightly differs from the one obtained from the effective medium approximation, but the results are close to the Monte-Carlo simulation results. Effects of ruggedness decrease with dimensionality. As expected, diffusion increases with dimension as the importance of the three site traps decreases rapidly with dimension.

Newman and Stein [18] discussed that diffusion in a rugged one-dimensional landscape is pathological due to repeated encounters with large barriers over a long period of time. A pathology arises in one dimension when an exchange of limits of large system size and long observation time leads to different results. This anomaly disappears as the dimension is increased to two because of the increase in escape routes with system size. This was found to be correct when diffusion in a system with a rugged energy landscape was studied in a computer simulation.[19] Thus, theoretical analyses allow us to connect the reduction of diffusion to accessible entropy.

## III. DIFFUSION-ENTROPY CORRELATION IN COLLOIDAL SYSTEMS

Several decades ago, Indrani and Ramaswamy explained an order of magnitude reduction in the long-time diffusion of colloidal particles suspended in a solution from the short-time diffusion.[20] While the short-time diffusion was determined by local viscosity, the long-time diffusion was dominated by additional frictional retardation due to long-range interaction between the colloidal particles. Through mode-coupling theory analysis, Indrani and Ramaswamy calculated the ratio of long-time diffusion to short-time diffusion as 0.05. Use of Eq.(1) (equating the short time relaxation times) would suggest that the reduction of entropy of the tagged colloids would arise from the establishment of correlations and approximately given by

$$\Delta S / k_B = \frac{3}{2} \ln \left( \frac{D}{D_S} \right) \simeq \frac{3}{2} \ln (0.05) = -4.49 \qquad (6)$$

This appears to be in the right range, although no detailed calculation to justify this has been attempted. This observation can be elucidated using the Sackur-Tetrode equation, commonly employed to estimate the entropy of an ideal monatomic gas.[21] According to the Sackur-Tetrode equation, entropy is significantly influenced by the particle mass. The mass of the colloid is approximately $10^3$-$10^4$ times the mass of argon.[22] This is possibly the reason why colloids exhibit much higher entropy change than that is expected for argon systems where change of entropy on freezing is about *1.3 $k_B$*, smaller than what is given by Eq.(6). We refer to the seminal work of de Schepper and Cohen on a kinetic theory of transport in dense interacting colloids [13], where the particles experience a cage effect for a long time, reducing diffusion and entropy.

As mentioned earlier, Eq.(6) suggests that one can obtain an estimate of the decrease of diffusion in terms of a decrease in the entropy of the colloidal system and vice versa. This could be an interesting proposition, which is also embodied in Rosenfeld scaling. However, there is a marked difference. The expression of Sorkin et al. potentially relates diffusion constants of two thermodynamic states, $D(S_1)$ and $D(S_2)$, to their entropy difference, and of course, it is useful if the relaxation times can be accurately specified. Rosenfeld scaling relates diffusion to excess entropy, which is the entropy difference between the entropy of the liquid and that of the ideal gas. However, we do recover a form close to Eq.(1) if we take the reference state as the ideal gas state, except that the *dimensionality factor d* would be missing in the Rosenfeld relation.

In a different study, Bhattacharyya et al. established a correlation between the entropy-based approach towards diffusion and the dynamical MCT approach in the sense that the change in structural correlation gives rise to change in entropy via Green's expression and also gives rise to slowing down in the MCT theory.[24]

**IV.     RANDOM WALK IN A LATTICE BY ESCAPE FROM A TRAP**

Diffusion in a lattice can be described as a random walk where diffusion (D) is related to the hopping rate (*k*) by *D= (1/2d) k $a^2$*, where *a* is the lattice spacing, and *d* denotes the dimension of the system. In a theoretical analysis that follows similar analyses by Zwanzig [25],

Acharya et al.[26] derived an approximate relation between diffusion and entropy in different ensembles, considering diffusion in a periodic potential with maxima and minima. The derivation considers a deterministic system. We start the proceedings with the rate expression $k(T)$ defined in the canonical ensemble for crossing a barrier, which reads as [27–30]

$$k(T) = \frac{1/2 \iint \frac{dq^{Nd} dp^{Nd}}{(2\pi\hbar)^{Nd}} \exp(-\beta H_N) \delta(q_1 - q_c)|\dot{q}_1|}{\iint \frac{dq^{Nd} dp^{Nd}}{(2\pi\hbar)^{Nd}} \exp(-\beta H_N)} \quad (7)$$

where $H_N$ is the total Hamiltonian of the system with $N$ degrees of freedom and is defined as $H_N = \sum_{i=1}^{N} \frac{\mathbf{p}_i^2}{2m} + V(\mathbf{q}_1,....\mathbf{q}_N)$. $\mathbf{p}_i$ denotes the conjugate momentum along the $i^{th}$ degrees of freedom of the particle of mass $m$, $d$ denotes the dimension of the system and $V(\mathbf{q}_1,....\mathbf{q}_N)$ is the total potential energy of the system. In the preceding equation $\dot{q}_1$ defines velocity along the reaction coordinate only. In the derivation of Eq.(7), we assume that one of the coordinates, i.e., reaction coordinate (say, $q_1$) is perpendicular to the dividing surface $\sigma(q_1 = q_c) = 0$, which separates the reactant from the product side where $q_c$ represents the critical value of $q_1$ at the transition point. We aim to get the simplified form of the numerator and denominator of Eq.(7). The denominator of Eq.(7) denotes the partition function in the canonical ensemble (Q). We now employ two well-known relations available in the canonical ensemble, i.e., $A = -k_B T \ln Q$ and $S = k_B \frac{\partial}{\partial T}(T \ln Q)$. Here, $A$ represents the Helmholtz free energy, $Q$ is the canonical partition function, $k_B$ is the Boltzmann constant, $T$ denotes temperature, and $S$ is entropy. We then simplify the denominator of Eq.(7) with the aid of the definition of entropy to obtain $Q = e^{-\frac{Nd}{2}} e^{\frac{S}{k_B}}$. We follow the same decomposition techniques of $H_N$ that we used in the microcanonical ensemble for simplifying the numerator and denominator of Eq.(7). After a short calculation, the numerator of Eq.(7) is simplified as

$$\frac{1}{\beta h}\exp(-\beta E_0)\frac{1}{(2\pi\hbar)^{Nd-1}}\iint dq^{Nd-1}dp^{Nd-1}\exp(-\beta H_{N-1}) \tag{8}$$

Now, we define $Q^\dagger$ as

$$Q^\dagger = \frac{1}{(2\pi\hbar)^{Nd-1}}\iint dq^{Nd-1}dp^{Nd-1}\exp(-\beta H_{N-1})$$
$$= e^{-\frac{Nd-1}{2}} e^{\frac{S^\dagger}{k_B}} \tag{9}$$

Where $S^\dagger$ is defined as $S^\dagger = k_B \frac{\partial}{\partial T}(T\ln Q^\dagger)$. Now, with the aid of Eq.(9) and Eq.(8), Eq.(7) becomes

$$k(T) = \left(\frac{k_B T}{h}\right)\exp(-\beta E_0)e^{1/2}\exp\left(-\frac{S-S^\dagger}{k_B}\right) \tag{10}$$

We can employ the relation between diffusion and escape rate in accordance with the random walk model to derive the exponential relation between diffusion and entropy. With a similar protocol, we can derive the *D-S* exponential relation in any ensemble. However, in this derivation, there is no explicit dimension *d in the exponent*. However, this is a transition state theory calculation, and there is no bare diffusion, as was the case in our discussions on diffusion in the rugged energy landscape. The dimensionality dependence in a Langevin equation-type description remains an open question, but see the discussion below.

In a recent study, Liao *et al.* used a projection operator technique to derive the relation between diffusion and entropy for a Brownian particle, which reads as, $D = \frac{\hbar}{eM}\exp[S/(k_B d)]$ .Interestingly, this relation resembles the Rosenfeld scaling relation, i.e., $D = a\ exp(bS/k_B)$ where $a = \dfrac{\hbar}{eM}$ and $b = \dfrac{1}{d}$.[31] Here, $M$ is the mass of the Brownian particle, $d$ denotes the dimensionality of the system and $\hbar$ is the Planck constant. However, this relation is different from the Rosenfeld scaling relation, where excess (over the ideal gas value) entropy is involved instead of the total entropy of the system. Nevertheless, it leads to the expression of Sorkin et al., except for a factor of two in the exponent, when we express diffusion in one entropy state in terms of diffusion in a different entropy state $D_1/D_2 = \exp\left[\dfrac{1}{k_B d}(S_1 - S_2)\right]$.

It is probably not surprising that while the activation energy term is independent of dimension, the entropy term exhibits a strong dimensionality dependence. In the rugged energy landscape picture, the availability of multiple paths of movement helps escape bottlenecks presented by maxima and minima.

## V. CROSSOVER BEHAVIOR IN DIFFUSION-ENTROPY RELATION

Rosenfeld-type exponential diffusion-entropy relations break down as the glass transition region is approached because while the value of the entropy decreases to a small value, the resultant decrease in diffusion (according to Rosenfeld and Sorkin) is predicted to be modest. *These relations cannot capture the sharp decline in the value of the diffusion.* One turns to the Adam-Gibbs type relation to explain the sharp decrease of diffusion as the entropy goes to small values. A clue to the crossover behavior comes from the theory of Indrani and Ramaswamy[20], which should be additionally applicable to the colloidal glass transition. As the interactions between colloids increase at larger volume fractions, the resultant caging leads to a lowering of long-time entropy that tends to zero. The point we want to stress here is that the long-time diffusion is decoupled from the short-time diffusion and short-time entropy and, therefore, can be described only by the long-time entropy that tends to zero.

A study by Bhattacharyya et al. could be relevant here [12,24]. In this study, a bridge between MCT[10,11] and RFOT[6–9] was proposed. The rate of relaxation of a dynamical quantity, like the stress-stress time correlation function, was written as a sum of the MCT and the RFOT contributions. An assumption introduced by Bhattacharyya et al. suggests the relaxation of the dynamic structure factor of a supercooled liquid rate could vary as $A\exp(aS_{ex}) + B\exp(b/S_c)$, where $s_C$ is the configuration entropy. The first term in the parenthesis of the exponent arises from the mode coupling theory-type analysis, while the second term arises from RFOT. In the detailed calculation that was undertaken by Bhattacharyya *et al.*, the effects of the two contributions to the relaxation function (like the density-density correlation function or the stress-stress time correlation function) *were determined self-consistently and were shown to provide a realistic description*. Nevertheless, such an analysis could be extended to explain the simultaneous occurrence of the two channels of relaxation mentioned above.

## VI.   CONCLUSIONS

One can understand the role of entropy in diffusion in a simple fashion. If we consider the time taken to make a displacement *δr* in a *d*-dimensional space, then the number of ways this can be achieved depends on phase space degeneracy. That is, the same magnitude displacement can be made in more ways in higher dimensions. This argument leads to Rosenfeld relation between diffusion and entropy.[32] However, this approach based on configuration space dynamics cannot address the two-step diffusion addressed by Sorkin *et al.*, Indrani and Ramaswamy, and Seki and Bagchi. One can establish a relation between short and long-time diffusion in all the latter discussions.

There are several unique aspects to the diffusion-entropy scaling problem. [1,14] Although the strong dependence of diffusion on entropy has long been recognized, its dimensionality dependence has not been widely discussed. The article by Sorkin et al. established a relationship that could be of practical value because it allows to express of diffusion in a different state in terms of the entropy difference between the two states.

However, despite a certain amount of success, as outlined above, no exact relation between diffusion and entropy appears to exist. The crossover scenario also remains ill-understood.

We conclude by pointing out a recent work that calculated Boltzmann's H-function, *H(t)* (which is sometimes regarded as the time-dependent analog of entropy), and showed that *H(t)* exhibits pronounced dependence on dimensionality. For a gas of Lennard-Jones spheres, *H(t)* decay was much slower in one dimension than in two and three dimensions.[33] The pathological behavior of diffusion in one dimension was manifested and captured even in Boltzmann's H-function.

## ACKNOWLEDGMENTS

We thank Professor G. Voth for pointing out the reference [1] to us. We also thank Professor K. Seki of AIST, Tsukuba, Japan and Mr. Shubham Kumar for his collaboration and discussions. This work is supported by support from the Indian Institute of Science to SA and by a DST-SERB India National Science Chair to BB.